\documentclass[12pt]{article}
\usepackage{amsmath}
\usepackage{amsfonts}
\usepackage{graphicx}
\begin{document}
\def\eg{{\em e.g.\,}}
\def\ie{{\em i.e.\,}}
\def\etc{{\em etc.\,}}
\def\C{\mathrm{C}}
\def\Q{\mathrm{Q}}
\def\R{\mathrm{R}}

\def\cH{\mathcal{H}}
\def\dof{{d.o.f.\,}}
\def\bra{\langle}
\def\ket{\rangle}
\def\av#1{\langle #1 \rangle}
\def\vi{\mathbf{i}}
\def\vk{\mathbf{k}}
\title{Quantum states of hierarchic systems}
\author{M.V.Altaisky\\ Joint Institute for Nuclear Research,
 Dubna, 141980, Russia; \\
 and Space Research Institute RAS, Profsoyuznaya 84/32, 
Moscow,\\ 117810, Russia, e-mail: altaisky@mx.iki.rssi.ru}
\date{Revised July 27}
\maketitle
\begin{abstract}
The density matrix formalism which is widely used in the theory 
of measurements, quantum computing, quantum description of 
chemical and biological systems always imply the averaging over 
all states of the environment. In practice 
this is impossible because the environment of the system is 
the complement of this system to the whole Universe and contains 
infinitely many degrees of freedom. A novel method of construction 
density matrix which implies the averaging only over the direct environment 
is proposed. The Hilbert space of 
state vectors for the hierarchic quantum systems is constructed.
\end{abstract}
\section{Introduction}
A progress in quantum computing, quantum chemistry and quantum 
description of biological systems ultimately requires mathematical 
methods for the description of quantum systems with hierarchic organization.
It is clearly impossible to account for each electron wave function in 
a living cell or a microprocessor. The methods of 
quantum statistical mechanics are of little use for the description 
of nonequilibrial systems. At the same time the problems of possible 
quantum superpositions in the systems of few and more atoms are now becoming  
practical problems of the creating of quantum computers \cite{nc2000}.  

If in quantum computing we need to create mesoscopic quantum superpositions,
in biology we need to explain the processes which take place in living cells 
and can be explained only at quantum level. 
Recent advances in biology show that functioning of living cells, 
first of all the information processing in brain \cite{HHT2002} and 
hereditary information 
processing in DNA replication \cite{patel2001} are essentially quantum: 
the superpositions of 
quantum states play important role in both the functioning of neurons in brain
and RNA translation on ribosome. Otherwise, the observed tremendous 
effectiveness of information processing in biological systems can not 
be explained. 

A key problem of quantum mechanics of mesoscopic objects is the 
problem of measurement \cite{mensky2000}. 
For a complex system consisting of a number 
of subsystems, we usually do not have direct access to the subsystems. 
A measurement of the angular momentum of a molecule leads to a reduction 
of the state vector to that with measured momentum, but the atoms 
in this molecule can be still in superposed states, and we may not have 
full control over the states of the subsystem affecting only the system as 
a whole. However we can get limited information on the subsystem measuring 
the state of the system as a whole, and we can partially control the states 
of the subsystem acting on the system. In this way if a measured projection 
of spin of a system of two spin-half fermions is 1, we are sure that both 
fermions have spin projection equal to $\frac{1}{2}$. To prepare a 
desired superposed 
state of the subsystem we often act by magnetic field on the system as 
a whole without causing full decoherence in the parts of the system. 
Similarly, if very big systems are considered, a medical treatment of 
certain organ in a living being is often done by changing the state of that 
being as a whole.     

In this paper we present a mathematical framework to describe the states 
of hierarchically organized systems, which do need quantum mechanical 
description.

\section{Quantum measurement}
A state of elementary quantum object can be represented as a normalized 
vector in Hilbert space 
$$
|\psi\ket = \sum_i c_i |\phi_i\ket, \quad \sum_i|c_i|^2=1 .$$
A measurement performed on a state $|\psi\ket$ with a probability 
$|c_i|^2$ reduced the superposition to either of the pure states 
$|\phi_i\ket$. This is von Neumann reduction of the wave function. 

Physically the measurement process takes place via a {\em measuring 
apparatus}, a system which should also obey the laws of quantum mechanics. 
So, to 
get the information about the state of a quantum system,
or to change its state in a prescribed way, we have 
to interact with combined system $S\!+\!B$, where ``B'' means buffer, or 
measuring apparatus. The measurement is understood as such interaction 
between the system $S$ and the apparatus  $B$ that changes the quantum state of 
$B$ in accordance to the state of $S$, \ie writes the information on $S$ into 
the state of $B$. 

Let the initial state of the measured system $S$ be 
\begin{equation}
|\psi\ket = c_1 |\phi_1\ket+c_2 |\phi_2\ket,
\label{psi2} 
\end{equation}
and let the initial state of the apparatus be $|\Phi_0\ket$. Let the 
apparatus $B$ respond to the initial states of the system $S$ by the 
following rule 
$$
\begin{array}{lcl}
\phi_1  &:& \Phi_0 \to \Phi_1 \\
\phi_2  &:& \Phi_0 \to \Phi_2 .
\end{array}
$$
Then the measurement process is a quantum transition 
\begin{equation}
\Psi = (c_1 |\phi_1\ket + c_2 |\phi_2\ket) |\Phi_0\ket \to 
\Psi'= c_1 |\phi_1\ket|\Phi_1\ket  + c_2 |\phi_2\ket|\Phi_2\ket.
\label{meas}
\end{equation}
The resulting state $|\Psi'\ket$ of the combined system $S\!+\!B$ is an 
{\em entangled} state, \ie it can not be factorized into a product of 
pure states of the system and the apparatus.  

The state of the 
combined system is described by the density matrix $\rho_{SB}$ rather 
than a state vector. To 
determine the density matrix of the system $S$ we have to trace $\rho_{SB}$ 
over the states of $B$
\begin{equation}
\rho_S = Tr_B(\rho_{SB}).
\end{equation}
Taking the trace over the $B$ states of the density matrix 
$\rho_{SB} = |\Psi'\ket\bra\Psi'|$, we get 
\begin{equation}
\rho_S = |c_1|^2 |\phi_1\ket\bra \phi_1| + |c_2|^2 |\phi_2\ket\bra \phi_2|
     + \bra \Phi_1|\Phi_2\ket c_2c_1^* |\phi_2\ket\bra \phi_1| 
     + \bra \Phi_2|\Phi_1\ket c_1c_2^* |\phi_1\ket\bra \phi_2|.
\label{rhomix}
\end{equation}    
As it is seen from \eqref{rhomix} the reduction of a quantum superposition 
to a mixture takes place only if the states of $B$ are mutually 
orthogonal: $\bra \Phi_2|\Phi_1\ket=0$. 
The state of the combined system $S+B$ can be then measured in the 
same way, by incorporating an extra device $B_1$ \etc. This goes 
to the end when the state of the final macroscopic device $B_N$ 
is measured classically. 

Usually, all measuring devices, buffers and so on which interact 
with quantum system are altogether understood as the {\em environment}.
If we measure a state of a single quantum system, the result of the 
measurement depends on both the state of the system and the state of the 
environment. The wave function of the composite system 
is 
\begin{equation}
|\psi\ket = \sum_{ij} C_{ij}|\phi_i\ket |\theta_j\ket,
\label{s_e}
\end{equation}
where $\bigl\{|\phi_i\ket\bigr\}_i$ is the complete set of the 
state vectors of the system, with $\bigl\{|\theta_i\ket\bigr\}_i$ 
being the complete set of the environment, \ie the rest of the 
Universe. Then, the equation \eqref{s_e} gives the most general expression 
for the wave function of ``system $\oplus$ environment''.  

For a variety of problems we need to know only the states and 
the evolution of the system in question.  
If it is the case, in a measurement performed on the system the wave 
function can be considered as a linear combination of different 
states of the system taken with weights dependent on the 
environment. This means the operators corresponding to the physical 
observables related to the system act only on $|\phi\ket$ vectors:
$$A |\phi_i\ket |\theta_j\ket = \bigl(A|\phi_i\ket \bigr)|\theta_j\ket$$
and the average of the observable $A$ is given by 
\begin{equation}
\av{A} \equiv \bra\psi|A|\psi\ket 
       = \sum_{ii'}\rho_{i'i} \bra\phi_i|A|\phi_{i'}\ket \equiv Tr(\rho A),
\end{equation}
where 
$$\rho_{i'i} \equiv \sum_j C_{ij}^* C_{i'j}$$ 
is the density matrix. (The orthonormality of the state vectors of 
the environment is assumed 
$
\bra \theta_i | \theta_j\ket = \delta_{ij}
$.)

The density matrix $\rho_{i'i}$, being Hermitian, is usually represented in 
the diagonal form 
$$\rho_{ii'} = \bra \phi_{i'}|\hat\rho|\phi_i \ket, 
\quad \hbox{where\ }  \hat\rho = \sum_i |i\ket\omega_i\bra i|,$$
where the eigenvalue $\omega_i$ is the probability of finding the 
system in the $i$-th eigenstate $Tr\,\rho=\sum_i\omega_i=1.$
Thus, instead of having the wave function of a microsystem ``which completely 
determines the state of that system'' we have a probabilistic description of 
the microsystem in terms of the density matrix $\rho$, derived by averaging 
over the states of the environment, usually understand as a macrosystem. 

The words {\em microsystem} and {\em macrosystem} are used here and after 
to denote the part and the whole, rather than quantum and classical. 
We assume the hierarchic organization of the matter. Each physical object 
can be characterized by the ladder of vicinities, or the entities which 
encompass it. A quark is inside a nucleon, a nucleon inside a nucleus and 
so on up to the scale of galaxies and the Universe itself. The words 
microsystem and macrosystem are used to describe the {\em partial ordering} in this 
ladder. If the system $A$ is a part of $B$, $A$ is called a microsystem 
with respect to $B$, and $B$ is called a macrosystem with respect to $A$.
  
\section{Wave function and density matrix of hierarchic system}
Geometrically we know that the microsystem is located {\em inside} 
the macrosystem. An electron is a part of atom, 
an atom is a part of a molecule \etc. This suggests, that instead of direct 
averaging over {\em all} degrees of freedom of the environment, to get the 
density matrix of the microsystem, we can represent the wave function of a 
microsystem in a hierarchic form, sequentially taking into account the 
states of all systems our system is a part of. 

\begin{figure}[th]
\centering \includegraphics[width=3in]{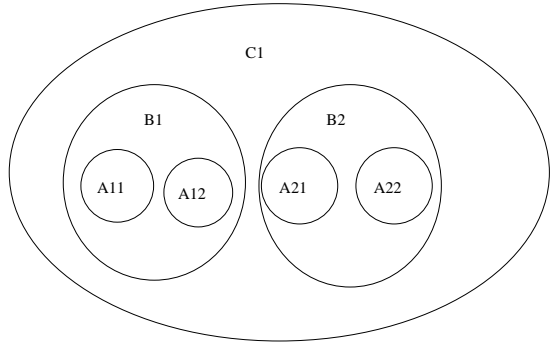}
\vspace*{8pt}
\caption{The structure of binary hierarchic system. 
The density matrix for the subsystem $A_{11}$ requires the 
averaging over the states of $B_1$ and $C_1$, but not over the states of 
$B_2,A_{21},A_{22}$.}
\label{hier1:pic}
\end{figure}

For instance, the wave function of the electronic system of an 
atom with $n$ electrons will be 
$$
\left\{ \psi_A, \left\{\psi_{Ae_1},\ldots,\psi_{Ae_n}, \right\} \right\},
$$
where $\psi_A$ is the wave function of the whole atom (a macrosystem), 
labeled by the total momentum $J$, the total orbital momentum $L$, and the 
total spin 
$S$ of the atom, \ie those depending not only on electron system, but 
also on nuclei. The wave functions of the electrons in atom $\psi_{Ae_k}$ 
are therefore different from the wave functions of free electrons 
$\psi_{e_k}$. Now we have an elegant way to get the information on quantum 
states of a subsystem by averaging over only a given branch of a hierarchy 
tree rather than all states of the environment. For instance, the 
state of a subsystem $A_{11}$, shown in Fig.~\ref{hier1:pic}, is given as a 
three component wave function $(\psi_{C_1},\psi_{C_1B_1},\psi_{C_1B_1A_{11}})$.
If required, the density matrix of such a system can be obtained by averaging 
over degrees of freedom of $C_1$ and $B_1$, but not the $B_2$. 

In general, to describe a state of an object 
$A_1$ (interacting with objects $A_2,\ldots,A_N$) which is a part of an object 
$B_1$ we have to write the wave function in the form 
\begin{equation}
\Psi = \{ \psi_{B_1},  \{ \psi_{B_1A_1},\ldots,\psi_{B_1A_N}\} \},
\label{eq1}
\end{equation}
where $\psi_{B_1}$ is the wave function of the whole (labeled by $B_1$), 
and $\psi_{B_1A_1}$ is the wave function of a {\sl 
component} $A_1$ belonging to the entity $B_1$.
For instance, $A_1,A_2,A_3$ may be quarks, and $B_1$ may be proton.
The objects $A_1,\ldots,A_N$ are {\em inside} $B_1$, and hence it is 
impossible to commute the operator-valued wave functions $[\Psi_{B_1},\Psi_{B_1A_1}]$ or 
to multiply them $\Psi_{B_1},\Psi_{B_1A_1}$. {\em The functions 
$\Psi_{B_1}(x)$ and  $\Psi_{B_1A_1}(x)$, taken in coordinate representation, 
live in different functional spaces}. To label the hierarchic object 
(\ie to set a coordinates on it) one needs a hierarchic tree, like those 
used in biology to trace the evolution. 
If the system $B$ consists of $k$ parts $A_1,\ldots A_k$, than we 
can write the wave function of the form \cite{bqm0007023}: 
\begin{equation}
|\psi\ket = \sum C^j_{i_1,\ldots,i_k}|\phi_{i_1}\ket\otimes\ldots\otimes
|\phi_{i_k}\ket |\theta_j\ket.
\label{s_i}
\end{equation}
The equation \eqref{s_i} is of course an approximation, neglecting the 
effects of the environment $U\setminus B$. It may be considered as a 
kind of mean-field approximation: all effects of the environment 
on the subsystems $A_k$ are taken into account only by means of the 
effect of $B$ to its subsystems. Further we shall call $\phi$ the 
{\em microlevel} and $\theta$ the {\em macrolevel} wave function.  

There are at least two aspects of the problem. First, if we know the 
eigenvectors $|\phi_{i_1}\ket,\ldots,|\phi_{i_k}\ket$ of all parts,  
the interaction between these parts and the effects of external fields, 
we should be able, in principle, to construct the set $|\theta_j\ket$. 
But the wave functions of all constituents can't be known simultaneously 
and it is more reasonable to introduce the state vector $|\theta_j\ket$ of 
the embracing system by hand, and consider the interaction between this 
the mean field  and the microlevel components. 

Second, it is well known in biology that the action of the external fields 
on the 
components of a cell strongly depends on the state of this cell as a whole. 
This can be said about radiation absorption etc. Thus there is a problem 
of control theory. How one can control the microlevel activity acting only \
on macrolevel. 

Let $A$ be an operator acting on the microlevel of a system containing 
$k$ subparts. Then 
\begin{equation}
\av{A} = \sum {C^*}^{j'}_{{i_1}',\ldots,{i_k}'}{C}^{j}_{{i_1},\ldots,{i_k}}
\bra\theta_{j'}| \bra \phi_{{i_1}'}|\ldots\bra\phi_{{i_k}'}|
A 
|\phi_{{i_1}}\ket,\ldots,|\phi_{{i_k}}\ket|\theta_j\ket   
      =\rho_{\vi,\vi'}\bra \vi |A|\vi'\ket,
\end{equation}
where $\vi \equiv (i_1,\ldots,i_k), 
|\vi\ket \equiv |\phi_{{i_1}}\ket,\ldots,|\phi_{{i_k}}\ket$ is the multiindex 
of the microlevel state.

If the operator $A=A_1$ acts only on the first ($i_1$) subsystem of 
the microlevel, the density matrix of this subsystem is obtained by the 
averaging over all other ($i_2,\ldots,i_k$) subsystems and the macrosystem 
state
\begin{equation}
\rho^{(1)}_{i_1 {i_1}'}= \sum_{j,i_2,\ldots,i_k} 
{C^*}^j_{{i_1}',i_2,\ldots,i_k}{C}^j_{{i_1},i_2,\ldots,i_k}.
\end{equation}
In analogy with the controlled gates in quantum computing algorithms, we can 
introduce operators which act on the microlevel depending on the state of the 
macrolevel \cite{bqm0110043}:
\begin{equation}
\hat B = |\vi\ket |\theta_m\ket B^m_{\vi\vk}\bra\theta_m|  \bra\vk|.
\end{equation}
The mean value of the corresponding observable in a two level hierarchic 
system is 
\begin{equation}
\av{B} = \bra\psi|\hat B|\psi\ket=\sum_{j,\vi,\vi'} {C^*}^j_\vi 
B^j_{\vi\vi'} C^j_{\vi'}.
\end{equation}

As an example, let us consider a system of two particles with spin 
$1/2$. The macrosystem of two such particles can be in either of three 
states: $S_z=0,+1,-1$. Thus a microsystem inside (say, a quark in meson) is 
in superposition of the states 
\begin{equation}
|\Psi\ket = c_{++} |\uparrow\ket |\uparrow\ket 
          + c_{+0} |\uparrow\ket |\rightarrow\ket
          + c_{--} |\downarrow\ket |\downarrow\ket 
          + c_{-0} |\downarrow\ket |\rightarrow\ket
\label{meson}.
\end{equation}
Since it is impossible for a system of spin $1$, to have $S_z=1$ having  
either of the component spins in the opposite direction, no other terms 
are present in the superposition \eqref{meson}. Taking into account 
the symmetry between up and down configurations, we get 
$$|c_{++}|=|c_{--}|=c_1, \quad |c_{+0}|=|c_{-0}|=c_0.$$
The density matrix for the subpart is 
\begin{equation}
\rho = \left( \begin{array}{cc}
               c_1^2 + c_0^2 &   c_0^2 \\
               c_0^2         &   c_1^2 + c_0^2
       \end{array} \right).
\end{equation}
The trace of the density matrix ${\rm} Tr(\rho) = 2(c_1^2 + c_0^2)=1$ 
corresponds to the normalization of the hierarchic state \eqref{meson} 
$\bra \Psi | \Psi \ket=1$.
Thus, measuring the state of the system by projection operator 
we get some information on its subsystems without acting on the 
subsystems wave functions. Say, if a meson is found to be in a state 
$|\uparrow\ket$, then by applying the projection operator 
$P_\uparrow = |\uparrow\ket\bra\uparrow|$ to its wave function, we get 
the information that both its constituent quark have the spin projections 
$s_z=\frac{1}{2}$; if meson has zero projection of spin, we can only say 
that its constituents have opposite projections of spins.  

\section{Pauli principle}
It is important, that the hierarchic representation of the wave function 
suggests a solution to the problem, whether or not two electrons 
belonging to macroscopically different objects can be in the same 
state. By construction, the hierarchic wave functions of  electrons 
in two different macrosystems carry the label of those macrosystems, 
and thus the states of the electrons are different being dependent on 
different labels. In fact, the solution of this problem was indicated 
by R.Peierls \cite{peierls}, who emphasized that the definition of the 
quantum state includes 
the coordinate, and since two electrons belong to different objects, they  
are in different states. 

More formally, we can assert that two fermions belonging to {\em the same 
entity of the next hierarchic level} can not have coinciding quantum 
numbers, but 
those belonging to different entities of the next hierarchic level 
can. For instance the systems $A_{11}$ and $A_{12}$ in 
Fig.~\ref{hier1:pic}, if being fermions, 
can not be in the same state; but at the same time $A_{11}$ can be in the same 
state as $A_{21}$. 

\section{The Hilbert space of hierarchic states}
The wave function components corresponding to different hierarchic levels 
may be of different 
nature: have different spin, isospin, color \etc and hence live in different 
spaces. The Hilbert space of hierarchic wave functions can be 
constructed by assuming common linearity at all hierarchy levels: 
$$\Psi_1,\Psi_2 \in \cH, a,b \in \C \Rightarrow 
\Psi = a\Psi_1+b\Psi_2 \in \cH.$$
If 
\begin{eqnarray*}
\Psi_1 &=& \{ \psi_{B_1},  \{ \psi_{B_1A_1},\ldots,\psi_{B_1A_N}\},\ldots \}, \\
\Psi_2 &=& \{ \psi_{D_1},  \{ \psi_{D_1C_1},\ldots,\psi_{D_1C_N}\},\ldots \},
\end{eqnarray*}
then their linear combination is 
\begin{equation}
\Psi = a\Psi_1+b\Psi_2 = \{ a\psi_{B_1}   +b\psi_{D_1}, 
                         \{ a\psi_{B_1A_1}+b\psi_{D_1C_1}, \ldots, 
                            a\psi_{B_1A_N}+b\psi_{D_1C_N}\}, \ldots\}
\label{multilin}.
\end{equation}
The scalar product is defined componentwise: 
\begin{equation}
\bra \Psi_1|\Psi_2\ket = \bra \psi_{B_1}|\psi_{D_1} \ket 
+ \sum_{i=1}^N \bra \psi_{B_1A_i}|\psi_{D_1C_i} \ket + \ldots.
\label{scprod}
\end{equation}
The norm of the vector in hierarchic space defined by scalar product 
is a sum of norms of all components: 
\begin{equation}
||\Psi||^2 = \bra \Psi_1|\Psi_1\ket = \bra \Psi_{B_1}|\Psi_{B_1}\ket 
                + \sum_{i=1}^N \bra \Psi_{B_1A_i}|\Psi_{B_1A_i}\ket 
                + \ldots.     
\label{mnorm}
\end{equation}

The second quantization procedure can be also defined in the spaces of 
hierarchic states in a straightforward way. If $B$ is a system which 
contains the subsystems $A_1,\ldots,A_N$, then the creation and annihilation 
operators act on the hierarchic states as follows
\begin{eqnarray*}
a^+(B)|0\ket = |B\ket, a(B)|B\ket = |0\ket, a(B)|0\ket = 0 |0\ket, \\
a^+(A_i) |B\ket = \{|B\ket,|BA_i\ket\},a(A_i)\{|B\ket,|BA_i\ket\}= |B\ket\\
a(B)|BA_i\ket = |A_i\ket.
\end{eqnarray*} 

Taking into account that a system is geometrically bigger 
then its subsystem, and therefore the step from subsystem to a system is a 
coarse-graining, we can see an analogy between the multiscale wave functions  
for an arbitrary quantum fields with the norm \eqref{mnorm}, and the 
decomposition of a scalar field with respect to representations of the 
affine group. Let $\phi(x)\in \mathrm{L}^2(\R^d)$ be a scalar field. 
Using its scale components 
$$\phi_a(b) := 
\int \frac{1}{a^{d/2}} \bar \psi\left(\frac{x-b}{a}\right)
\phi(x) d^dx,$$ 
the field $\phi$ can be represented in a form 
of decomposition with respect to the affine group $g:x\to ax+b$:
\begin{equation}
\phi(x) = \frac{1}{C_\psi} \int \frac{1}{a^{d/2}} 
\psi\left(
\frac{x-b}{a}
\right)
\phi_a(b) \frac{dad^db}{a^{d+1}},
\end{equation} 
where $C_\psi$ is a normalization constant, which depends on the 
basic function $\psi$ only. The scalar product and the norm of the vector 
can be taken in either of representations: 
\begin{equation}
\int |\phi(x)|^2 dx = \int |\phi_a(b)|^2 \frac{dad^db}{a^{d+1}}.
\label{wnorm}
\end{equation}
The equation \eqref{wnorm} is apparently a continuous 
counterpart of the equation \eqref{mnorm} in case when all hierarchic 
components are scalar fields.  

\section{On sad fate of the Schr\"odinger Cat}
The Schr\"odinger cats, as well as all other cats, are quantum systems  
with tremendous number of degrees of freedom. That's why according to  
quantum mechanics the life time of any coherent superposition 
of such big systems is very short. In hierarchic description presented 
above the possibility of a superposition 
$$\frac{|\hbox{``cat alive''}\ket + |\hbox{``cat decayed into parts''}\ket}{\sqrt2}
$$
takes a surprising form. A hierarchic description wave function of 
an alive cat looks like 
\begin{equation}
\begin{array}{lcl}
\Psi_{\hbox{alive cat}} &=& \Bigl\{ \psi_{cat}, 
\left\{ \psi_{cat\to head},\psi_{cat\to pams},\ldots,\psi_{cat\to tail}
\right\},\ldots,\\ 
& &\{\psi_{cat\to head \to eye \to \ldots\to e_-},\ldots \},
\ldots,
\Bigr\}
\end{array}.
\label{alivecat}
\end{equation}
In contrast, a dead cat as just ``a collection of parts of the cat'' 
(the words taken from the letter from Einstein to Heisenberg, 1950), so 
there should be no first term $\psi_{cat}$ (describing the entity 
``cat as a whole'') in the description of a dead head. The hierarchic 
wave function of a dead cat will look like 
\begin{equation}
\Psi_{\hbox{dead cat}} = \Bigl\{  
\left\{ \psi_{head},\psi_{pams},\ldots,\psi_{tail}
\right\},\ldots,\{\psi_{head \to eye \to \ldots\to e_-},\ldots \},
\ldots,
\Bigr\}.
\label{deadcat}
\end{equation}

The hierarchic wave function of an alive cat \eqref{alivecat} and that of 
dead cat \eqref{deadcat} can not be in a superposition because they 
have different structure. Clearly there is now interference in the 
first term $\psi_{cat} + nothing$, and unlikely there is an interference 
in the next terms. Say, $\psi_{cat\to head}$ wave function in \eqref{alivecat} 
may have different structure and much less components than $\psi_{head}$ in 
\eqref{deadcat}, for some of the information used to construct the head of an alive cat may have been taken from $\psi_{cat}$.

Concerning the other living beings, we dare to say that there a few characters 
of the living matter, which are not present in non-living matter. Namely:
\begin{enumerate}
\item   The properties of a living system are more than just a collection 
        of its component properties. In other words, it is impossible to 
        predict the whole set of properties of a complex biological system 
        even having known all properties of its components and their 
        interactions.
\item   The properties and functions of the components of a system depend 
        on the state of the whole system. In other words, the same components 
        being included in different systems may have different properties.
\item   For a non-living matter, at least in principle, we can calculate the 
        wave function of a big molecule by multiplying the wave functions of 
        all the electrons and nucleons in this molecule using 
        the Clebsh-Gordon coefficients. For a living system we can not do 
        that, in a sense that the result of such multiplication will not 
        give an adequate description of the system.
\end{enumerate}
\section{Conclusion}
The understanding of physics of life and consciousness requires new 
mathematical methods applicable to complex systems far from equilibrium. 
One of the perspective approaches is the application of quantum information 
theory methods to biological system. Needless to say that the information 
theory 
itself often works as an ultimate tool to describe biological systems, 
where nonequilibrial state and strong interaction with environment 
precludes the application of standard quantum mechanics and thermodynamics. 

In its turn, the study of biological systems by quantum mechanics and 
quantum information methods may be expected to yield a technical solution 
of the problem of long living coherent states in many-particle systems, which 
are so required for the construction of quantum computers, but are very 
likely to be present in brain \cite{HHT2002}. The hierarchic organization of all living 
systems may be the key to the problem of preserving the many particle systems  
in a coherent superposition safe from the environmental decoherence even 
at room temperatures. 

A mathematical framework for the description of hierarchic quantum states 
is presented in this paper and the previous 
papers \cite{bqm0007023,bqm0110043}.
 
\section*{Acknowledgement}
The author is thankful to Dr. B.F.Kostenko for useful comments.
The work was partially supported by Russian Foundation for Basic Research, 
Project 03-01-00657.

\end{document}